\documentclass[12pt]{article}
\usepackage{stmaryrd}
\usepackage[dvips]{graphicx}
\usepackage{amssymb,amsmath,amsfonts,palatino,amsthm}
\usepackage{amssymb,mathtools}
\usepackage[sans]{dsfont}
\usepackage{booktabs}
\newcommand{\beq}{\begin{equation}}
\newcommand{\eeq}{\end{equation}}
\newcommand{\beqq}{\begin{equation*}}
\newcommand{\eeqq}{\end{equation*}}

\newcommand{\mc}{\mathcal}
\newtheorem{theorem}{Theorem}

\newtheorem{condition}{Condition}
\newtheorem{definition}{Definition}
\begin{document}

\title{C, P, and T of Braid Excitations in Quantum Gravity}
\author{Song He\thanks{Email address:
hesong@pku.edu.cn}\\ \\School of Physics, Peking University,
Beijing, 100871, China, \\ \\ \\Yidun Wan\thanks{Email address:
ywan@perimeterinstitute.ca}
\\ \\
Perimeter Institute for Theoretical Physics,\\
31 Caroline st. N., Waterloo, Ontario N2L 2Y5, Canada, and \\
Department of Physics, University of Waterloo,\\
Waterloo, Ontario N2J 2W9, Canada\\}
\date{February 15, 2008}
\maketitle
\vfill
\begin{abstract}
We study the discrete transformations of four-valent braid
excitations of framed spin networks embedded in a topological
three-manifold. We show that four-valent braids allow seven and only
seven discrete transformations. These transformations can be
uniquely mapped to C, P, T, and their products. Each CPT multiplet
of actively-interacting braids is found to be uniquely characterized
by a non-negative integer. Finally, braid interactions turn out to
be invariant under C, P, and T.
\end{abstract}
\vfill
\newpage
\tableofcontents
\newpage

\section{Introduction}
Since a ribbonized preon model\cite{Bilson-Thompson2005} was coded
into local braided ribbon excitations\cite{Bilson-Thompson2006}
there has been a large amount of research effort towards a quantum
theory of gravity with matter as topological
invariants\cite{Hackett2007, Wan2007, LeeWan2007, HackettWan2008,
LouNumber, HeWan2008b, Isabeau2008}. However, a serious limitation
of the results of \cite{Bilson-Thompson2006} was realized soon that
the conservation laws which preserve the braid excitations are
exact. In other words, there is no possibility of dynamics of these
exactly conserved excitations, e.g. creation and
annihilation\cite{Hackett2007}. Consequently, interpreting braid
excitations found in \cite{Bilson-Thompson2006} as particles - in
particular the Standard Model particles - is not going to work out
unless interactions are successfully introduced to that model.

Meanwhile, a new model has also been put forward, which solves the
problem of interaction and opens new interesting directions worth of
investigation\cite{Wan2007, LeeWan2007}. The new model involves
framed four-valent rather than three-valent spin networks, embedded
in a topological three-manifold, which also gives rise to local
braid excitations, each of which is a 3-strand braid formed by the
three common edges of two adjacent nodes of the network, and bases
the dynamics on the dual Pachner moves naturally associated with
four-valent graphs. The two main reasons of this extension are: that
four-valent graphs and the corresponding dual Pachner moves
naturally occur in spin foam models\cite{spin-foam}, and that
vertices of four-valent spin networks have true correspondence to
three-dimensional space.

The stable three-strand braids, under certain stability condition,
are local excitations\cite{LeeWan2007, Isabeau2008}. Among all
stable braids, there is a small class of braids which are able to
propagate on the spin network. The propagation of these braids are
chiral, in the sense that some braids can only propagate to their
left with respect to the local subgraph containing the braids, while
some only propagate to their right and some do both\cite{Wan2007,
LeeWan2007}. There is another small class of braids, the
actively-interacting braids; each is two-way propagating and is able
to merge with its neighboring braid when the interaction condition
is met\cite{LeeWan2007}. Braids that are not propagating are
christened stationary braids.

\cite{Wan2007, LeeWan2007} are based on a graphic calculus developed
therein. However, although the graphic calculus has its own
advantages - in particular in describing, e.g. the full procedure of
the propagation a braid, it is not very convenient for finding
conserved quantities of a braid which are useful to characterize the
braid as a matter-like local excitation. In view of this,
\cite{HackettWan2008, HeWan2008b} proposed an algebraic notation of
our braids and derived conserved quantities by means of the new
notation.

One of our goals is to see whether some braid excitations of
embedded 4-valent spin networks can eventually correspond to the
standard model particles or are more fundamental matter degrees of
freedom. Because CPT is a symmetry of quantum field theories, in
this paper we investigate the discrete transformations of 3-strand
braids of embedded 4-valent spin networks and map them to C, P, and
T transformations and their products. We will see that the
interaction of braids defined in \cite{LeeWan2007} respects CPT.

In fact, as a follow-up work of \cite{Bilson-Thompson2006}, in
\cite{LouNumber} a similar study of CPT-symmetry is being taken for
three-valent spin networks. However, in the 3-valent case there is
no dynamics, it is then unlikely to define what C, P, and T are
meant to be physically. Besides, the largest discrete symmetry in
the 3-valent case is $S_3\times \mathbb{Z}_2$, giving more than C,
P, T, and their products. On the contrary, in the 4-valent case we
have dynamics which surprisingly and strongly constraints the number
of possible discrete transformations to be exactly seven, excluding
the identity, which are allowed on 3-strand braids. Moreover, the
algebra developed and conserved quantities of braids found in our
companion paper\cite{HeWan2008b} also helps finding and mapping
discrete transformations of braids to C, P, and T. This will become
clear soon in the
sequel. Let us summarize the main results of this paper as follows:%
\begin{enumerate}
\item Discrete transformations C, P, T, and their products are found
on 3-strand braids of embedded framed 4-valent spin networks.
\item Reversing the momentum direction of a braid is understood to be
unambiguously associated with the flipping of the propagation
chirality of the braid.
\item The "electric" charge of a braid is naturally represented by
the effective twist number of the braid. A braid's spin is argued to
be related to the spin network labels on the braid.
\item Each CPT multiplet of actively-interacting braids is uniquely characterized by a non-negative integer.
\item Interactions of braids are found to be invariant under C, P, and
T separately, and is thus invariant under CPT.
\item Possible future developments by means of tensor categories are pointed
out.
\end{enumerate}
\section{Notation}
\begin{figure}
[h]
\begin{center}
\includegraphics[
height=2.1715in, width=2.8193in
]%
{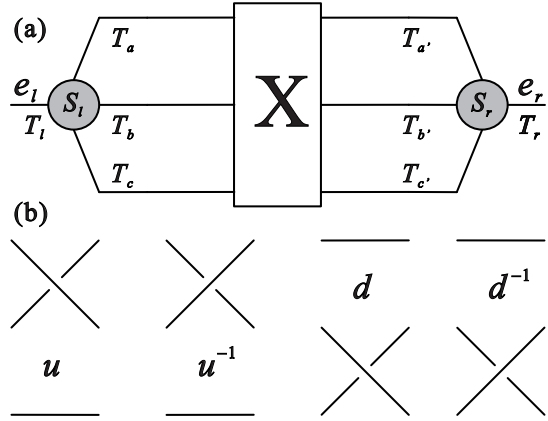}%
\caption{(a) is a generic 3-strand braid diagram formed by the three
common edges of two end-nodes. $S_l$ and $S_r$ are the states of the
left and right \textbf{end-node}s respectively, taking values in $+$
or $-$. $X$ represents a sequence of crossings, from left to right,
formed by the three strands between the two nodes. $(T_a, T_b,T_c)$
is the triple of the \textbf{internal twists} respectively on the
three strands from top to bottom, on the left of $X$. $T_l$ and
$T_r$, called \textbf{external twists}, are respectively on the two
external edges $e_l$ and $e_r$. All twists are valued in
$\mathbb{Z}$ in units of $\pi/3$\cite{Wan2007}. (b) shows the four
generators of $X$, which are also generators of the braid group
$B_3$.}%
\label{braid}%
\end{center}
\end{figure}

It is worth of re-emphasizing an essential point. A 3-strand braid
is a local sub network of the whole framed spin network embedded in
a topological 3-manifold; however, many embeddings are diffeomorphic
to each other, which gives rise to diffeomorphic (also called
equivalent in our approach) braids. We study a braid through its
2-dimensional projection, called a \textbf{braid diagram}. We
therefore will not distinguish braids from braid diagrams unless an
ambiguity arises. A generic example of such a braid diagram is
depicted in Fig. \ref{braid}(a). Equivalent braid diagrams form an
equivalence class. To choose an efficient representative of an
equivalence class of braids is important; in \cite{LeeWan2007} where
we studied propagation and interactions of braids our choice was to
represent an equivalence class by the representative which is a
braid diagram which has zero external twists, which simplifies the
interaction condition and the graphic calculus developed in
\cite{Wan2007, LeeWan2007}. Each class has one and only one such
representative. Thus a braid represented this way is said to be in
its unique representation.

An algebraic notation and the corresponding calculational method of
braids were put forward in \cite{HackettWan2008}, which applies only
to actively-interacting braids. This notation is extended to account
for propagating braids and even stationary braids\cite{HeWan2008b}.
Nevertheless, in this paper we first analyze the discrete
transformations of braids which will appear to be more transparent
and lucid in terms of the graphic calculus in some cases. On the
other hand, to identify the discrete transformations with C, P, T
and their products and to sort out the conserved quantities the
algebraic notation is more efficient. Therefore, we will use both
the algebraic notation and the graphic notation.

Let us briefly recall the algebraic notation of a braid introduced
in \cite{HeWan2008b}. A generic braid shown in Fig. \ref{braid}(a)
is characterized by an 8-tuple: $\{T_l,S_l,T_a,T_b,T_c,X,S_r,T_r\}$.
The crossing sequence $X$ satisfies the definition of an ordinary
3-strand braid, an element of the braid group $B_3$; hence it is
generated by the four generators shown in Fig. \ref{braid}(b). The
generators are assigned integral values according to their
handedness, namely $u=d=1$ and $u^{-1}=d^{-1}=-1$. Therefore,
crossings in the $X$ of a braid can also be summed over to obtain an
integer, the so-called \textbf{crossing number}:
$\sum_{i=1}^{|X|}x_i$, of the braid, where $|X|$ is the number of
crossings forming $X$ and $x_i$ is a crossing in $X$.

The $X$ of a braid induces a permutation $\sigma_X$, which is an
element in the permutation group $S_3$, of the three strands of the
braid. The triple of internal twists on the left of $X$ and the one
of the right of $X$ are thus related by
$(T_a,T_b,T_c)\sigma_X=(T_{a'},T_{b'},T_{c'})$ and
$(T_a,T_b,T_c)=\sigma^{-1}_X(T_{a'},T_{b'},T_{c'})$. That is,
$\sigma_X$ is a left-acting function of the triple of internal
twists, while its inverse, $\sigma^{-1}_X$ is a right-acting
function. The inverse relation between $\sigma_X$ and
$\sigma^{-1}_X$ is understood as:
\begin{equation}
\sigma^{-1}_X\left((T_a,T_b,T_c)\sigma_X\right)=\left(\sigma^{-1}_X(T_a,T_b,T_c)\right)\sigma_X\equiv(T_a,T_b,T_c)
\label{eqSigmaTsigma}
\end{equation}
Note that the twists such as $T_a$ and $T_{a'}$ are abstract
and have no meaning until their values and positions in a triple are
fixed. Thus, $(T_a,T_b,T_c)=(T'_a,T'_b,T'_c)$ means $T_a=T'_a$, etc,
and $-(T_a,T_b,T_c)=(-T_a,-T_b,-T_c)$. A generic braid diagram can
now be denoted concisely by
$$\left._{T_l}^{S_l}\hspace{-0.5mm}[(T_a,T_b,T_c)\sigma_X]^{S_r}_{T_r}\right.,$$
or by
$$\left._{T_l}^{S_l}\hspace{-0.5mm}[\sigma^{-1}_X(T_{a'},T_{b'},T_{c'})]^{S_r}_{T_r}\right..$$

Since an active braid is always equivalent to trivial braid diagrams
with external twists\cite{Wan2007}, and is usually represented by
one such trivial braid\cite{HackettWan2008, HeWan2008b}, it can be
denoted as
$$\left._{T_l}^{\hspace{0.75mm}S}\hspace{-0.5mm}[T_a,T_b,T_c]^{S}_{T_r}\right.,$$
where both end-nodes are in the same state. This is called a trivial
representation. However, a non actively-interacting braid is more
conveniently to be represented by its unique representative with
zero external twist, namely
$$\left._{\hspace{1mm}0}^{S_l}\hspace{-0.5mm}[(T_a,T_b,T_c)\sigma_X]^{S_r}_{0}\right..$$
Because a main purpose of this paper is to find discrete
transformations on all braids, we therefore will write an arbitrary
braid in its generic form in most of the paper. The unique
representation and trivial representation of braids will also be
used when it is good to do so.

\section{Discrete transformations}
Though not separately, as a theorem the combined action of the three
discrete transformations C, P and T, namely CPT, is a symmetry in
any Lorentz invariant, local field theory. Being a concrete model of
QFT, the Standard Model respects the CPT-symmetry too. The 3-strand
braids of embedded 4-valent spin networks are local excitations,
continuous transformations such as the equivalence moves of which
have been analyzed in \cite{Wan2007, HackettWan2008, HeWan2008b}, it
is then natural to look for the possible discrete transformations of
these local excitations and check their correspondence with C, P,
and T transformations. If the braids in our model would eventually
be mapped to the Standard Model particles, or even if they are more
fundamental entities on their own, which do not directly correspond
to the Standard Model particles, they should be characterized by
quantum numbers which have certain properties under the
transformations of C, P, and T. In fact, investigating the action of
discrete transformations on our braid excitations can help us to
construct quantum numbers of a braid, such as spin, charge, etc, out
of the characterizing 8-tuple of the braid. If the 8-tuple, which
only contains topological information of the embedding and framing,
is not sufficient to produce all necessary quantum numbers, we may
have to take spin network labels into account. One will see that
this is indeed the case.

Due to the dynamics of the braids of embedded 4-valent spin
networks, namely the propagation and interactions, there exist very
natural constraints on the discrete transformations one can apply.
The reason is that we should not allow that, for example, a discrete
transformation turns an actively interacting braid into one which is
not because there is no such a transformation in QFT which magically
changes a particle to something else, and vice versa. Similarly, we
can obtain other necessary rules. As a guideline, all rules are
listed below as a condition.

\begin{condition}\label{condLegalDT}A legal discrete transformation $\mc{D}$ on an arbitrary
braid $B$ must meet:
\begin{enumerate}
\item If $B$ is actively-interacting, then $\mc{D}(B)$ also actively
interacts.
\item If $B$ is not actively-interacting, $\mc{D}(B)$ must
remain so.
\item If $B$ is one-way (two-way) propagating, $\mc{D}(B)$ must still be one-way (two-way) propagating;
however, the propagation chirality of $B$ may be reversed in the
one-way case.
\item If $B$ is stationary, $\mc{D}(B)$ is stationary as well.
\end{enumerate}
\end{condition}

\subsection{The group of discrete transformations}
It is more convenient to write all discrete transformations in a
compact, algebraic form. This can be achieved by introducing the
so-called \textbf{atomic discrete operations} acting on the crossing
sequence, the end-nodes, the triples of internal twists, and the
pair of external twists separately. Each atomic transformations is
not qualified as a legal discrete transformation on its own due to
the violation of Condition \ref{condLegalDT}. However, a legal
discrete transformation can be written as a unique combination of
the atomic ones. All atomic transformations are defined and listed
with sufficient details in Appendix I, we thus in the rest of the
main text will directly use them without further explanation but
only a reference to the definition of each of them upon its first
appearance.

In view of the parity transformation in QFT, the first kind of
discrete transformations one may come up with is the mirror imaging
of a braid. However, there are two ways of mirror reflecting a
braid: one is to have the mirror perpendicular to the plane on which
the braid is projected, the other is to arrange the mirror parallel
to and behind the plane. Let us study them in order.
\begin{figure}
[ht]
\begin{center}
\includegraphics[
height=1.8948in, width=2.8193in
]%
{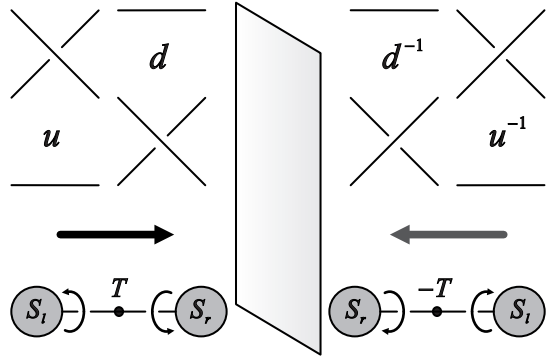}%
\caption{Shows how the crossing generators, end-nodes, twists, and
the propagation chirality (indicated by the two thick arrows) of a
braid (on the left) are mapped to their mirror images (on the right)
via a mirror \emph{perpendicular}
to the plane on which the braid is projected.}%
\label{mirrorPerp}
\end{center}
\end{figure}

Fig. \ref{mirrorPerp} illustrates the former case. Although only two
generators of the crossing sequence $X$ of an arbitrary braid is
shown in this figure, it is not hard to see that the order of the
crossings of the original $X$ on the left of the mirror must be
reversed by the mirror, resulting an $\mc{R}(X)$ (Def. \ref{defR})
of the mirror image of the original braid. Besides, every crossing
in $X$ is inverted by the mirror, giving rise to an $\mc{I}_X$ (Def.
\ref{defIx}). As a result, the mirror imaging takes the $X$ to
$X^{-1}$. In Fig. \ref{mirrorPerp} uses only one edge between the
two end-nodes of a braid to demonstrate the sign change of the twist
of the edge by the mirror. This is sufficient to show that all
twists of a braid should have a sign change via mirror imaging
because the sign of a twist is unambiguously defined everywhere of
an embedded spin network\cite{Wan2007}. This means that the atomic
operation $\mc{I}_T$ (Def. \ref{defIt}) must be part of this mirror
image transformation. One can also find from Fig. \ref{mirrorPerp}
that due to the exchange of the two end-nodes, which indicates the
atomic operations $\mc{E}_S$ (Def. \ref{defEs}) and $\mc{E}_{T_e}$
(Def. \ref{defEte}), and the existence of an $\mc{R}$, the left
triple of internal twists should be exchanged with the right triple
of internal twists, i.e. an $\mc{E}_T$ (Def. \ref{defEt}) is
involved. These observations provide us an explicit definition of
this mirror imaging as follows.
\begin{definition}
The \textbf{perpendicular mirror imaging} is such a discrete
transformation, denoted by $\mc{M}_{\perp}$, that
$$\mc{M}_{\perp}=\mc{E}_S\mc{E}_{T_e}\mc{E}_T\mc{I}_T\mc{I}_X\mc{R},$$ and that for a
generic braid
$B=\left._{T_l}^{S_l}\hspace{-0.5mm}[(T_a,T_b,T_c)\sigma_X]^{S_r}_{T_r}\right.$,
with $(T_a,T_b,T_c)\sigma_X=(T_{a'},T_{b'},T_{c'})$
\begin{equation}
\mc{M}_{\perp}(B)=\left._{-T_l}^{\ \ S_r}\hspace{-0.5mm}[-(T_{a'},T_{b'},T_{c'})\sigma_{X^{-1}}]^{S_l}_{-T_r}\right.,%
\end{equation}
with $-(T_{a'},T_{b'},T_{c'})\sigma_{X^{-1}}=-(T_a,T_b,T_c)$.
\end{definition}
A important point to address is that $\mc{M}_{\perp}$ flips the
propagation chirality. That is, if a braid $B$ is (left-)
right-propagating, then $\mc{M}_{\perp}(B)$ is (right-)
left-propagating, which is readily seen from Fig. \ref{mirrorPerp}
because a braid propagating towards the mirror from the left is
mirrored to a braid towards the mirror from the right. This surely
has no impact on a two-way propagating braid. Since $\mc{M}_{\perp}$
is simply a mirror image, an actively-interacting braid stays so
under this discrete transformation. Therefore, $\mc{M}_{\perp}$
fulfills Condition \ref{condLegalDT} and hence is indeed a legal
discrete transformation of braids. Concrete graphic examples are
shown in Appendix II.

\begin{figure}
[h]
\begin{center}
\includegraphics[
height=1.3482in, width=2.8046in
]%
{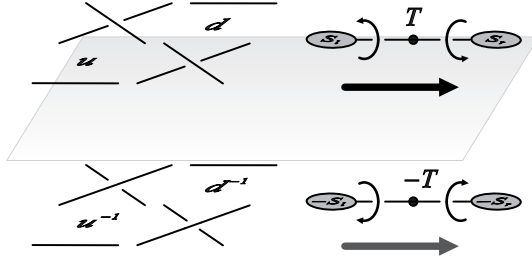}%
\caption{The crossing generators, end-nodes, twists, and the
propagation chirality of a braid (above the mirror) are mapped to
their mirror images (below the mirror) via a mirror \emph{parallel}
to the plane on which the braid is projected.}%
\label{mirrorPara}
\end{center}
\end{figure}

Fig. \ref{mirrorPara} presents the second type of mirror imaging of
a braid, where the mirror is parallel to and beneath the plane on
which the braid is projected. In contrast to the mirror imaging of
the first kind, the second kind does not reverse the order of the
crossings and does not exchange the two end-nodes, which leads to no
exchange of triples of internal twists. However, from Fig.
\ref{mirrorPara} it is clear that all crossings, twists, and the two
end-node states are inverted, resulting in three atomic operations:
$\mc{I}_X$, $\mc{I}_T$, and $\mc{I}_S$. As implied by the two thick
arrows respectively above and below the mirror in Fig.
\ref{mirrorPara}, the propagation chirality of a braid should not be
reversed under this mirror imaging. This implies that a stationary
braid remains stationary under this transformation. It is not hard
to see that this type of mirror imaging of an actively-interacting
braid must still be active. Therefore, we have another legal
discrete transformation of braids, as defined below.
\begin{definition}
The parallel mirror imaging, $\mc{M}_{\Box}$, is a discrete
transformation in the form
$$\mc{M}_{\Box}=\mc{I}_X\mc{I}_T\mc{I}_S,$$ such that for a generic
braid,
$B=\left._{T_l}^{S_l}\hspace{-0.5mm}[(T_a,T_b,T_c)\sigma_X]^{S_r}_{T_r}\right.$,
with $(T_a,T_b,T_c)\sigma_X=(T_{a'},T_{b'},T_{c'})$
\begin{equation}
\mc{M}_{\Box}(B) = \left._{-T_l}^{\ \
\bar{S_l}}\hspace{-0.5mm}[-(T_a,T_b,T_c)\sigma_{\mc{I}_X(X)}]^{\bar{S_r}}_{-T_r}\right.,
\end{equation}
with $-(T_a,T_b,T_c)\sigma_X=-(T_{a'},T_{b'},T_{c'})$.
\end{definition}

\begin{figure}
[h]
\begin{center}
\includegraphics[
height=2.3134in, width=3.0312in
]%
{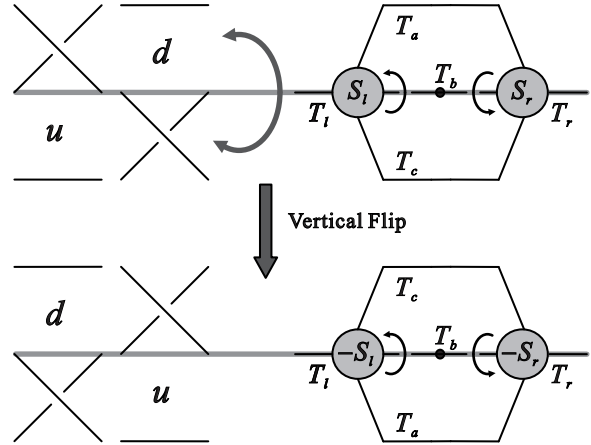}%
\caption{The discrete transformation, the \textbf{vertical flip}, as
a flip of a braid diagram along the axis (the thick grey line)
overlapped with the central strand of the braid. The transformation
of crossing
generators are shown on the left part.}%
\label{Fv}
\end{center}
\end{figure}
After talking about reflections, it is now the turn to study other
possibilities. The first is called a \textbf{vertical flip}, is
depicted in Fig. \ref{Fv}, in which, rather than showing the flip of
a whole generic braid with respect to the axis (the thick grey
horizontal line in the figure), we illustrate how the generators of
a crossing sequence and a trivial braid diagram without crossings
are transformed under such a flip, by which one can easily determine
the corresponding transformation of an arbitrary braid. Note that
this flip is not an equivalence rotational move, which is
continuous, defined in \cite{Wan2007} but rather a discrete
operation, taking a braid, e.g. the one in the upper part of Fig.
\ref{Fv}, directly to the one in the lower part of the same figure,
without any continuous intermediate steps; hence, no extra twists or
crossings are created or annihilated.

According to Fig. \ref{Fv}, the vertical flip neither reverses the
order of crossings nor exchange the two end-nodes of a braid;
however, it turns an upper crossing into a lower one and a lower one
to an upper one, with their handedness unchanged, which gives rise
to an $\mc{S}_c$ (Def. \ref{defSc}), the chain shift of the crossing
sequence $X$. From the figure an $\mc{I}_S$ is also obtained. An
interesting property of the vertical flip is that it swaps the top
and bottom internal twists of a braid, as seen in Fig. \ref{Fv},
leading to an $\mc{S}_T$ (Def. \ref{defSt}). The last atomic
operation involved in this rotation is seen from the figure to be an
$\mc{I}_S$.

A braid's propagation chirality is left intact under the vertical
flip because its reducibility is unchanged due to the fact that,
although each of it's end-node states is flipped, the crossing next
to each end-node is shifted to its counterpart of the same
handedness, which ensures a reducible end-node being again reducible
after the transformation. A stationary braid is thus still
stationary under this transformation. By the same argument, a braid,
which is actively-interacting, remains so too after the
transformation. Therefore, the vertical flip exhibited in Fig.
\ref{Fv} is indeed a legal discrete transformation. We now present
its explicit definition.
\begin{definition}
The \textbf{vertical flip}, $\mc{F}_V$, is a discrete transformation
in the form $$\mc{F}_V=\mc{I}_S\mc{S}_T\mc{S}_c,$$ which for a
generic braid
$B=\left._{T_l}^{S_l}\hspace{-0.5mm}[(T_a,T_b,T_c)\sigma_X]^{S_r}_{T_r}\right.$,
with $(T_a,T_b,T_c)\sigma_X=(T_{a'},T_{b'},T_{c'})$, satisfies%
\begin{equation}
\mc{F}_V(B)=\left._{T_l}^{\bar{S_l}}\hspace{-0.5mm}[(T_c,T_b,T_a)\sigma_{\mc{S}_c(X)}]^{\bar{S_r}}_{T_r}\right.,
\end{equation}
with $(T_c,T_b,T_a)\sigma_{\mc{S}_c(X)}=(T_{c'},T_{b'},T_{a'})$.
\end{definition}

One may try to find out all other legal discrete transformations in
a similar way. Nevertheless, our study shows that fortunately the
aforementioned three discrete transformations and their products,
seven altogether are the only allowable ones. In other words,
$\mc{M}_{\perp}$, $\mc{M}_{\Box}$, and $\mc{F}_V$ generate the
largest group of legal discrete transformations which, denoted by
$G_\mc{D}$, contains eight elements, including the identity
transformation. This group of discrete transformations and their
action on a generic braid is recorded in Table \ref{tabDT}.
\begin{table}[h]
\begin{center}%
\begin{tabular}
[c]{c c c c}
\toprule
Discrete  & Algebraic & Action on  & Prop-\\
Transformation & Form & $B=\left.  _{T_{l}}^{S_{l}}\hspace{-0.5mm}[(T_{a}%
,T_{b},T_{c})\sigma_{X}]_{T_{r}}^{S_{r}}\right.  $ &
Chirality\\[0.75ex]\midrule
$\mathds{1}$ & $\mathds{1}$ & $\left.  _{T_{l}}^{S_{l}}\hspace{-0.5mm}%
[(T_{a},T_{b},T_{c})\sigma_{X}]_{T_{r}}^{S_{r}}\right.  $ &
$+$\\[0.75ex]\midrule
$\mc{M}_{\perp}$ & $\mc{E}_{S}\mc{E}_{T_{e}}\mc{E}_{T}\mc{R}\mc{I}_{X}%
\mc{I}_{T}$ & $\left.  _{-T_{r}}^{\ \ S_{r}}\hspace{-0.5mm}[-(T_{a^{\prime}%
},T_{b^{\prime}},T_{c^{\prime}})\sigma_{X^{-1}}]_{-T_{l}}^{S_{l}}\right.
$ & $-$\\[0.75ex]\midrule
$\mc{M}_{\Box}$ & $\mc{I}_{S}\mc{I}_{X}\mc{I}_{T}$ & $\left.  _{-T_{l}%
}^{\ \ \bar{S_{l}}}\hspace{-0.5mm}[-(T_{a},T_{b},T_{c})\sigma_{\mc{I}_{X}(X)}]_{-T_{r}%
}^{\bar{S_{r}}}\right.  $ & $+$\\[0.75ex]\midrule
$\mc{F}_{V}$ & $\mc{I}_{S}\mc{S}_{c}\mc{S}_{T}$ & $\left.  _{T_{l}}^{\bar{S_{l}}%
}\hspace{-0.5mm}[(T_{c},T_{b},T_{a})\sigma_{\mc{S}_{c}(X)}]_{T_{r}}^{\bar{S_{r}}%
}\right.  $ & $+$\\[0.75ex]\midrule
$\mc{F}_{H}=\mc{M}_{\perp}\mc{M}_{\Box}$ & $\mc{I}_{S}\mc{E}_{S}\mc{E}_{T_{e}%
}\mc{E}_{T}\mc{R}$ & $\left.  _{T_{r}}^{\bar{S_{r}}}\hspace{-0.5mm}[(T_{a^{\prime}%
},T_{b^{\prime}},T_{c^{\prime}})\sigma_{\mc{R}(X)}]_{T_{l}}^{\bar{S_{l}}}\right.
$ & $-$\\[0.75ex]\midrule
$\mc{M}_{\perp}\mc{F}_{V}$ & $\mc{I}_{S}\mc{E}_{S}\mc{I}_{T}\mc{E}_{T_{e}%
}\mc{E}_{T}\mc{S}_{T}\mc{S}_{c}\mc{I}_{X}\mc{R}$ & $\left.  _{-T_{r}}^{\ \ \bar{S_{r}}%
}\hspace{-0.5mm}[-(T_{c^{\prime}},T_{b^{\prime}},T_{a^{\prime}})\sigma
_{\mc{I}_{X}\mc{S}_{c}\mc{R}(X)}]_{-T_{l}}^{\bar{S_{l}}}\right.  $ &
$-$\\[0.75ex]\midrule
$\mc{M}_{\Box}\mc{F}_{V}$ &
$\mc{I}_{T}\mc{S}_{T}\mc{S}_{c}\mc{I}_{X}$ & $\left. _{-T_{l}}^{\ \
S_{l}}\hspace{-0.5mm}[-(T_{c},T_{b},T_{a})\sigma
_{\mc{I}_{X}\mc{S}_{c}(X)}]_{-T_{r}}^{S_{r}}\right.  $ &
$+$\\\midrule
$\mc{F}_{V}\mc{F}_{H}$ & $\mc{E}_{S}\mc{S}_{T}\mc{E}_{T_{e}}\mc{E}_{T}%
\mc{S}_{c}\mc{R}$ & $\left. _{T_{r}}^{S_{r}}\hspace{-0.5mm}[(T_{c^{\prime}%
},T_{b^{\prime}},T_{a^{\prime}})\sigma_{\mc{S}_{c}\mc{R}(X)}]_{T_{l}}^{S_{l}%
}\right.  $ & $-$\\\bottomrule
\end{tabular}
\caption{The group $G_\mc{D}$ and its action on a generic braid
diagram. The last column shows whether the propagation chirality of
a braid changes under the corresponding transformations in the first
column; a $-$ means flipped and a $+$ means unaffected.}
\label{tabDT}
\end{center}
\end{table}

One can easily check that $G_\mc{D}$, consisting of the
transformations shown in the first column of Table \ref{tabDT}, is
indeed a group. That $G_\mc{D}$ is the largest group of legal
discrete transformations of 3-strand braids is a result of the fact
that $G_\mc{D}$ exhausts all possible combinations of the atomic
operations defined in Appendix I, which meet Condition
\ref{condLegalDT}, and that there exists no new atomic discrete
operations which can be constructed and combined with the current
ones without violating Condition \ref{condLegalDT}.

\subsection{Conserved quantities}
It is seen from Table \ref{tabDT} that each discrete transformation
takes a braid to a new braid which is not equivalent to the original
though they may be the same in some special cases. Focusing on the
third column of Table \ref{tabDT}, one readily finds that some
characterizing quantities are invariant under some discrete
transformations but are not under others. There are also composite
quantities constructed from the characterizing 8-tuple, which are
conserved under certain and changed under other transformations. It
is then necessary and helpful to explicitly demonstrate these
quantities, which is achieved by Table \ref{tabDTcq}, not only for
the purpose of mapping the discrete transformations to C, P, and T
but also for the task to sort out the physically meaningful quantum
numbers of a braid.
\begin{table}[h]
\begin{center}
\bigskip%
\begin{tabular}
[c]{ccccccccc}%
\toprule%
$G_{\mc{D}}$ & $\left(  S_{l},S_{r}\right)  $ & $\left(
T_{l},T_{r}\right)  $
& $[T_{a},T_{b},T_{c}]$ & $\sum\limits_{i=a}^{c}T_{i}$ & $\sum\limits_{i=1}%
^{|X|}x_{i}$ & $\Theta^{\prime}$ &
$T_{l}+T_{r}$ & $\Theta$\\\midrule%
$\mc{M}_{\perp}$ & $\left( S_{r},S_{l}\right)  $ & $-\left(
T_{r},T_{l}\right)  $ &
$-[T_{a^{\prime}},T_{b^{\prime}},T_{c^{\prime}}]$ &
$-$ & $-$ & $-$ & $-$ & $-$\\\midrule%
$\mc{M}_{\Box}$ & $\left(  \bar{S}_{l},\bar{S}_{r}\right)  $ &
$-\left(
T_{l},T_{r}\right)  $ & $-[T_{a},T_{b},T_{c}]$ & $-$ & $-$ & $-$ & $-$ & $-$\\\midrule%
$\mc{F}_{V}$ & $\left(  \bar{S}_{l},\bar{S}_{r}\right)  $ & $\left(
T_{l},T_{r}\right)  $ & $[T_{c},T_{b},T_{a}]$ & $+$ & $+$ & $+$ & $+$ & $+$\\\midrule%
$\mc{F}_{H}$ & $\left(  \bar{S}_{r},\bar{S}_{l}\right)  $ & $\left(
T_{r},T_{l}\right)  $ &
$[T_{a^{\prime}},T_{b^{\prime}},T_{c^{\prime}}]$ & $+$
& $+$ & $+$ & $+$ & $+$\\\midrule%
$\mc{M}_{\perp}\mc{F}_{V}$ & $\left(  \bar{S}_{r},\bar{S}_{l}\right)
$ &
$-\left(  T_{r},T_{l}\right)  $ & $-[T_{c^{\prime}},T_{b^{\prime}%
},T_{a^{\prime}}]$ & $-$ & $-$ & $-$ & $-$ & $-$\\\midrule%
$\mc{M}_{\Box}\mc{F}_{V}$ & $\left(  S_{l},S_{r}\right)  $ &
$-\left(
T_{l},T_{r}\right)  $ & $-[T_{c},T_{b},T_{a}]$ & $-$ & $-$ & $-$ & $-$ & $-$\\\midrule%
$\mc{F}_{V}\mc{F}_{H}$ & $\left(  S_{r},S_{l}\right)  $ & $\left(  T_{r}%
,T_{l}\right)  $ & $[T_{c^{\prime}},T_{b^{\prime}},T_{a^{\prime}}]$
& $+$ &
$+$ & $+$ & $+$ & $+$%
\\\bottomrule
\end{tabular}
\caption{Conserved quantities of a generic braid diagram under
discrete transformations in $G_{\mc{D}}$. $|X|$ is the number of
crossings.
$\Theta=\sum\limits_{i=a}^{c}T_i+T_l+T_r-2\sum\limits_{i=1}^{|X|}x_i$
is the \textbf{effective twist}. $\Theta ' =\Theta - (T_l+T_r)$ is
called the \textbf{internal effective twist}.}%
\label{tabDTcq}
\end{center}
\end{table}
The reason to consider the quantities listed in Table \ref{tabDTcq},
apart from their properties under discrete transformations, is their
conservation under interactions. The quantity $\Theta$, defined in
\cite{HeWan2008b} as the \textbf{effective twist number}, of a
single braid is conserved under both equivalence moves and evolution
moves\cite{Wan2007, LeeWan2007}. It is also an additive conserved
quantity under interactions of two braids, in the sense that the
$\Theta$-value of the resulted braid of the interaction of two
braids is equal to the sum of the $\Theta$-values of the two braids
before the interaction\cite{LeeWan2007}. This conservation law,
though obtained in representing a braid by its unique
representative, is independent of the choice of the representative
of the braid.

In \cite{HackettWan2008, HeWan2008b}, however, we also studied
representing a braid by an extremum of it, i.e. an equivalent braid
diagram with the least number of crossings (defined in
\cite{Wan2007}). An actively-interacting braid has infinite number
of extrema, namely the trivial braid diagrams with external twists,
which are the very trivial representative aforementioned.
Fortunately, it is shown in \cite{HackettWan2008} that all extrema
of an actively-interacting braid share the same value of the sum of
the two external twists, i.e. $T_l+T_r$. Likewise, a propagating
braid also has infinite number of extrema; however, as conjectured
in \cite{HeWan2008b}, all extrema of a propagating braid have the
same $T_l+T_r$, $\sum\limits_{i=a}^{c}T_i$, and
$\sum\limits_{i=1}^{|X|}x_i$ as well. So we may define another
quantity, the \textbf{internal effective twist}, $\Theta
'=\Theta-(T_l+T_r)$, which is the same for all extrema of a braid
and is equal to $\Theta$ when the equivalence class of braid
diagrams is represented by the unique representative free of
external twists.

Importantly, \cite{HackettWan2008, HeWan2008b} showed that both
$T_l+T_r$ and $\Theta '$, and hence $\Theta$ are additive conserved
quantities under interactions. More precisely, for two braids, $B_1$
with $\Theta'_1$ and $B_2$ with $\Theta'_2$, the resulted braid
$B_1+B_2$ has $\Theta'=\Theta_1+\Theta_2$.

In addition, all discrete transformations affect the end-nodes of a
braid. However, the only meaningful quantity, made of the end-node
states of a braid, is the so-called effective state,
$\chi=S_lS_r(-)^{|X|}$, which is a conserved quantity of braids
under equivalence moves and also is a multiplicative conserved value
under interactions of braids\cite{HeWan2008b}. Table \ref{tabDTcq}
reads that none of the discrete transformations changes the number
of crossings of a braid; therefore, all discrete transformations
preserve the $\chi$ of a braid.

\section{C, P, and T}
Since a braid is a local excitation, regardless of whether it
corresponds to a Standard Model particle or not one can at least
make an analogy between it and a one-particle state. This is the
main task of this section.%
\subsection{Finding C, P, and T}
To map the discrete transformations found in the last section to C,
P, T, and their products, it is helpful to recall how the latter
ones act on single particle states in the context of quantum field
theory.
\begin{table}[h]
\begin{center}%
\begin{tabular}
[c]{ll}%
\toprule%
& $\left\vert \mathbf{p},\sigma,n\right\rangle $\\\midrule%
$\mc{C}$ & $\propto\left\vert \mathbf{p},\sigma,n^{c}\right\rangle $\\\midrule%
$\mc{P}$ & $\propto\left\vert -\mathbf{p},\sigma,n\right\rangle $\\\midrule%
$\mc{T}$ & $\propto(-)^{J-\sigma}\left\vert
-\mathbf{p},-\sigma,n\right\rangle$\\\midrule%
$\mc{CP}$ & $\propto\left\vert -\mathbf{p},\sigma,n^{c}\right\rangle $\\\midrule%
$\mc{CT}$ & $\propto(-)^{J-\sigma}\left\vert -\mathbf{p},-\sigma
,n^{c}\right\rangle $\\\midrule%
$\mc{PT}$ & $\propto(-)^{J-\sigma}\left\vert
\mathbf{p},-\sigma,n\right\rangle$\\\midrule%
$\mc{CPT}$ & $\propto(-)^{J-\sigma}\left\vert \mathbf{p},-\sigma
,n^{c}\right\rangle $%
\\\bottomrule%
\end{tabular}%
\caption{The action of $\mc{C}$, $\mc{P}$, $\mc{T}$, and their
products on a one-particle state, where $\mathbf{p}$ is the
3-momentum, $\sigma$ is the third component of the particle spin
$J$, and $n$ stands for the charge.}%
\label{tabCPT}
\end{center}
\end{table}
In Table \ref{tabCPT}, we chose to denote C, P, and T
transformations in the Hilbert space by calligraphic letters
$\mc{C}$, $\mc{P}$, and $\mc{T}$. For this reason we have already
used calligraphic letters for the legal discrete transformations of
braids as well because braids are local excitations of embedded spin
networks which are the states in the Hilbert space describing the
fundamental space-time.

We emphasize here three things. Firstly, so far we have not
incorporated spin network labels, which are normally representations
of gauge groups, and that our scheme in this section is to obtain
the map between two groups of transformations mentioned above by
trying to utilize topological characterizing quantities of a braid
as much as possible, without involving spin network labels.
Secondly, for now we do not take into account the phase and sign
factors in Table \ref{tabCPT}. Finally, all the transformations are
restricted to local braid states, rather than a full evolution
picture. By doing so, surprisingly, there is a unique such map.
Nevertheless, these two issues will be discussed in the next
section.

According to Table \ref{tabCPT}, the four transformations $\mc{P}$,
$\mc{T}$, $\mc{CP}$, and $\mc{CT}$ reverse the three momentum of a
one-particle state. But then what do we mean by the momentum of a
braid? In the case of Loop Quantum Gravity, there has not been a
well-defined Hamiltonian yet but a Hamiltonian constraint which does
not assign a well-defined energy and hence neither a momentum to a
local excitation. In fact, the issue is more fundamental in the
sense that what do we mean by a direction in space when there is no
a notion of fundamental space-time but only superposed spin networks
which might lead to a (semi-) classical space-time under some
continuous limit? In the case of spin networks as a concept of
quantum geometry in general \cite{Penrose}, this problem of
direction has not been solved either.

Nonetheless, we do not need an explicitly defined 3-momentum of a
braid to pick out the discrete transformations which can flip the
braid's momentum. Each braid has a propagation chirality, namely it
is either left-propagating or right-propagating or both. Propagation
chirality is a locally defined property of a braid with respect to
its neighboring subgraph which can be projected horizontally on the
plane one is looking at. Actually the propagation chirality is an
intrinsic property of a braid and is not the same as the propagation
direction of the braid; the latter should be viewed with respect to
the whole spin network embedded in a topological 3-manifold.
Consequently a braid can actually propagate in any "direction" with
respect to its spin network or to an observer, regardless its
propagation chirality and how the semiclassical geometry is
obtained. One may imagine looking at a braid which moves on a spin
network along a "circle" and comes back to its original location.

However, locally, i.e. within a sufficiently small subgraph
containing a braid, the braid's propagation chirality coincides with
its propagating direction. An immediate result of this is that if
the local propagation chirality of a braid is flipped by a discrete
transformation, so is its propagation direction. The direction of
the 3-momentum of braid, by any means it is defined, is associated
with the propagation direction. Therefore, the discrete
transformations reversing the 3-momentum of a braid are exactly
those flipping the propagation chirality of the braid, which are,
according to Table \ref{tabDT}, $\mc{M}_{\perp}$, $\mc{F}_H$,
$\mc{M}_{\perp}\mc{F}_V$, and $\mc{F}_H\mc{F}_V$.

In other words, $\mc{M}_{\perp}$, $\mc{F}_H$,
$\mc{M}_{\perp}\mc{F}_V$, and $\mc{F}_H\mc{F}_V$ are the only legal
discrete transformations which can possibly be identified with
$\mc{P}$, $\mc{T}$, $\mc{CP}$, and $\mc{CT}$, and our task is to
find precisely which is which. We know that $\mc{P}$ is the
transformation which does not change any quantum number but the
3-momentum of a particle. Hence the discrete transformation of a
braid which reasonably corresponds to a $\mc{P}$ must have the
fewest effects on the braid. From Tables \ref{tabDT} and
\ref{tabDTcq} one can see that transformations $\mc{F}_H$ and
$\mc{F}_V\mc{F}_H$ are the two candidates because they both reverse
the 3-momentum without negating the twists, crossing values, and
hence effective twists. Furthermore, $\mc{F}_H$ exchanges the left
and the right triples of internal twists, but on top of this,
$\mc{F}_V\mc{F}_H$ swaps the first and the third twists in the
triple of internal twists of a braid. Therefore, $\mc{F}_H$ is the
only candidate of a $\mc{P}$.

We need two more correspondences to pin down the complete mapping.
For this we should find quantum numbers of a braid which are, or
analogous to, charge and spin. Among all the conserved quantities of
a braid, composed of characterizing topological quantities of the
braid, only the total effective twist number $\Theta$ is independent
of the representative of the equivalence class of the braid. Other
conserved quantities, e.g. $\Theta '$, are not. We know that
charges, e.g. electric and color, are unambiguous quantum numbers a
particle. Consequently, representative-dependent conserved
quantities of a braid, though maybe useful in other ways, should not
be considered as charges. This means only $\Theta$ can be a
candidate of determining certain charges of a braid.

Moreover, there are actually two more reasons, which are more
heuristic and physical. As we know, the electric charge of a
particle is quantized to be multiples of $1/3$. Now the interesting
thing is that all our twists and hence the effective twists happen
to be integers in units of $1/3$\footnote{A twist was said to be in
units of $\pi/3$ previously in this paper. However, the $\pi$ here
only means half of a full rotation of a node with respect to one of
its edges, which can thus be normalized to one\cite{Wan2007}.} too;
the $1/3$ arises naturally rather than being put in by
hand\cite{Wan2007}. This is also an advantage of the 4-valent case
because in the 3-valent case, in contrast, a factor of $1/3$ must be
set by hand\cite{LouNumber}. On the other hand, the framing of our
spin networks which takes an edge to a tube is in fact a $U(1)$
framing; a tube coming from the framing of an edge is essentially an
isomorphism from $U(1)$ to $U(1)$. If a tube is twist free, it
simply means an identity map, whereas a twisted tube represents a
non-trivial isomorphism. That is to say, a twist can be thought as
characterizing the isomorphisms on $U(1)$ spaces. An interesting
fact is that the electric charge is due to a $U(1)$ gauge symmetry.
These suggest that $\Theta$ or an appropriate function of it can be
viewed as the "electric charge" of a braid, which might serve as an
explanation of why electric charge is quantized so.

Bearing this in mind, Table \ref{tabDTcq} presents four discrete
transformations, $\mc{M}_{\perp}$, $\mc{M}_{\Box}$,
$\mc{M}_{\perp}\mc{F}_V$, and $\mc{M}_{\Box}\mc{F}_V$, which negate
the $\Theta$ value of a braid and hence correspond to $\mc{CP}$,
$\mc{C}$, $\mc{CT}$, and $\mc{T}$ in certain manner. Our strategy is
to find $\mc{C}$ first. Since a $\mc{C}$ does not flip the momentum,
so does a $\mc{CPT}$, the transformations $\mc{M}_{\Box}$ and
$\mc{M}_{\Box}\mc{F}_V$ are candidates of $\mc{C}$ and $\mc{CPT}$
for that they preserve the momentum, whereas $\mc{M}_{\perp}$ and
$\mc{M}_{\perp}\mc{F}_V$ are possibly $\mc{CP}$ and $\mc{CT}$.

On a single particle state, a $\mc{CPT}$ has one more effect than a
$\mc{C}$ because it also turns $\sigma$, the $z$-component spin, to
$-\sigma$. We notice that a $\mc{M}_{\Box}\mc{F}_V$ affects a braids
more than a $\mc{M}_{\Box}$ does; it swaps the first and the third
elements in the triple of internal twists of a braid besides adding
a negative sign to $\Theta$. As a result, although we do not know
what of a braid behaves like the $\sigma$, we can now consider the
transformation $\mc{M}_{\Box}$ as $\mc{C}$ and the transformation
$\mc{M}_{\Box}\mc{F}_V$ to be $\mc{CPT}$.

Given the three correspondences we now have, it is easy to track
down all the rest. As a summary, we list the map between
$G_{\mc{D}}$ and the group generated by $\mc{C}$, $\mc{P}$, $\mc{T}$
in Table \ref{tabMap}.
\begin{table}
\begin{center}
\begin{tabular}
[c]{ccccccc}%
\toprule%
$\mc{C}$ & $\mc{P}$ & $\mc{T}$ & $\mc{CP}$ & $\mc{CT}$ & $\mc{PT}$ &
$\mc{CPT}$\\\midrule%
$\mc{M}_{\Box}$ & $\mc{F}_{H}$ &
$\mc{F}_{V}\mc{F}_{H}$ & $\mc{M}_{\perp}$ &
$\mc{M}_{\perp}\mc{F}_{V}$ & $\mc{F}_{V}$ & $\mc{M}_{\Box}\mc{F}_{V}$%
\\\bottomrule
\end{tabular}%
\caption{The map between legal discrete transformations of braids
and $\mc{C}$, $\mc{P}$, $\mc{T}$, and their products.}%
\label{tabMap}
\end{center}
\end{table}
\subsection{CPT multiplets of braids}
With the C, P, and T we have found, one can see that certain
diffeomorphism-inequivalent braids may not be totally different from
each other, in the sense that they can belong to the same CPT
multipilet. it would be very interesting to see if a CPT multiplet
of braids has any characteristic property. It turns out that only
actively-interacting braids which belong to a CPT multiplet have a
clear and unique common topological property. We would like to
formulate this claim as the theorem below.
\begin{theorem}
Each CPT multiplet of actively-interacting braids is uniquely
characterized by a non-negative integer, $k$ - the number of
crossings of all braids in the multiplet, when they are represented
in the unique representation.
\end{theorem}
\begin{proof}
This theorem has a two-fold meaning: that all actively-interacting
braids with the same number of crossings belong to the same CPT
multiplet, and that all actively-interacting braids in a CPT
multiplet must have the same number of crossings. We prove the
former first. As demonstrated in \cite{LeeWan2007,Wan2007}, a braid
can interact actively if and only if it is completely reducible from
both ends, and its end-nodes are in opposite states for odd number
of crossings, and in the same state for even number of crossings.
Bearing this in mind and by \cite{Wan2007}, we can straightforwardly
work out the forms of all actively-interacting braids in the unique
representation with $k$ crossings for $k$ even
\begin{eqnarray}
B_1&=&^{\hspace{0.75mm}+}\hspace{-0.5mm}[(T_{1a},T_{1b},T_{1c})\sigma_{
(ud)^{k/2}}]^{+}\\\nonumber
B_2&=&^{\hspace{0.75mm}+}\hspace{-0.5mm}[(T_{2a},T_{2b},T_{2c})\sigma_{
(ud)^{-k/2}}]^{+}\\\nonumber
B_3&=&^{\hspace{0.75mm}-}\hspace{-0.5mm}[(T_{3a},T_{3b},T_{3c})\sigma_{
(du)^{k/2}}]^{-}\\\nonumber
B_4&=&^{\hspace{0.75mm}-}\hspace{-0.5mm}[(T_{4a},T_{4b},T_{4c})\sigma_{
(du)^{-k/2}}]^{-},\label{eqActiveEven}
\end{eqnarray}
and for $k$ odd,
\begin{eqnarray}
B'_1&=&^{\hspace{0.75mm}+}\hspace{-0.5mm}[(T'_{1a},T'_{1b},T'_{1c})\sigma_{
d(ud)^{(k-1)/2}}]^{-}\\\nonumber
B'_2&=&^{\hspace{0.75mm}+}\hspace{-0.5mm}[(T'_{2a},T'_{2b},T'_{2c})\sigma_{
d(ud)^{-(k+1)/2}}]^{-}\\\nonumber
B'_3&=&^{\hspace{0.75mm}-}\hspace{-0.5mm}[(T'_{3a},T'_{3b},T'_{3c})\sigma_{
u(du)^{(k-1)/2}}]^{+}\\\nonumber
B'_4&=&^{\hspace{0.75mm}-}\hspace{-0.5mm}[(T'_{4a},T'_{4b},T'_{4c})\sigma_{
u(du)^{-(k+1)/2}}]^{+},\label{eqActiveOdd}
\end{eqnarray}
where we have omitted all the external twists for that they are zero
in the unique representation. The triple of internal twists of each
actively-interacting braid in the unique representation is uniquely
determined by the crossing sequence and end-node states of the braid
for it to interact actively\cite{LeeWan2007}. In addition, if an
exponent in Eq. \ref{eqActiveEven} or \ref{eqActiveOdd} is positive,
it means, for example, $(ud)^2=udud$. By adopting from Appendix I
the definition of $X^{-1}$ with respect to $X$, the meaning of the
negative exponents in Eqs. \ref{eqActiveEven} and \ref{eqActiveOdd}
is clear: for instance, $(ud)^{-2}=d^{-1}u^{-1}d^{-1}u^{-1}$.

It is straightforward to see that if we apply $\mc{M}_{\Box}$,
$\mc{F}_V$, and $\mc{M}_{\Box}\mc{F}_V$, or according to
Table\ref{tabCPT}, $\mc{C}$, $\mc{P}\mc{T}$ and $\mc{C}\mc{P}\mc{T}$
on $B_1$ for even $k$'s and $B'_1$ for odd $k$'s, we get
\begin{eqnarray}
\mc{C}(B_1)&=&^{\hspace{0.75mm}-}\hspace{-0.5mm}[-(T_{1a},T_{1b},T_{1c})\sigma_{
(du)^{-k/2}}]^{-}\\\nonumber
\mc{PT}(B_1)&=&^{\hspace{0.75mm}-}\hspace{-0.5mm}[(T_{1c},T_{1b},T_{1a})\sigma_{
(du)^{k/2}}]^{-}\\\nonumber
\mc{CPT}(B_1)&=&^{\hspace{0.75mm}+}\hspace{-0.5mm}[-(T_{1c},T_{1b},T_{1a})\sigma_{
(ud)^{-k/2}}]^{+}\label{eqCPTActiveEven}
\end{eqnarray}
for even $k$'s, and
\begin{eqnarray}
\mc{C}(B'_1)&=&^{\hspace{0.75mm}-}\hspace{-0.5mm}[-(T'_{1a},T'_{1b},T'_{1c})\sigma_{
u(du)^{-(k+1)/2}}]^{+}\\\nonumber
\mc{PT}(B'_1)&=&^{\hspace{0.75mm}-}\hspace{-0.5mm}[(T'_{1c},T'_{1b},T'_{1a})\sigma_{
u(du)^{(k-1)/2}}]^{+}\\\nonumber
\mc{CPT}(B'_1)&=&^{\hspace{0.75mm}+}\hspace{-0.5mm}[-(T'_{1c},T'_{1b},T'_{1a})\sigma_{
d(ud)^{-(k+1)/2}}]^{-}\label{eqCPTActiveOdd}
\end{eqnarray}
for odd $k$'s. Comparing Eq. \ref{eqActiveEven} with Eq.
\ref{eqCPTActiveEven}, and Eq. \ref{eqActiveOdd} with Eq.
\ref{eqCPTActiveOdd}, one can see that the crossing sequence and end
node states of $\mc{C}(B_1)$, $\mc{PT}(B_1)$ and $\mc{CPT}(B_1)$ are
exactly the same as that of $B_4$, $B_3$ and $B_2$ respectively, for
even $k$'s; similar observation holds for odd $k$'s as well. As for
the internal twists, since they are uniquely determined by crossing
sequence and end node states, one must have
\begin{eqnarray*}
(T_{2a},T_{2b},T_{2c})&=&-(T_{1c},T_{1b},T_{1a})\\
(T_{3a},T_{3b},T_{3c})&=&(T_{1c},T_{1b},T_{1a})\\
(T_{4a},T_{4b},T_{4c})&=&-(T_{1a},T_{1b},T_{1c})
\end{eqnarray*}
for braids with even crossings, and similar relations for braids
with odd crossings. Therefore, in the unique representation, all
actively-interacting braids with $k$ crossings are related to each
other by discrete transformations as following,
\begin{eqnarray}
\mc{C}(B_1)&=&B_4\\\nonumber \mc{PT}(B_1)&=&B_3\\\nonumber
\mc{CPT}(B_1)&=&B_2
\end{eqnarray}
for even $k$'s, and
\begin{eqnarray}
\mc{C}(B'_1)&=&B'_4\\\nonumber \mc{PT}(B'_1)&=&B'_3\\\nonumber
\mc{CPT}(B'_1)&=&B'_2
\end{eqnarray}
for odd $k$'s.

Pointed out in the last section, none of the discrete
transformations on a braid diagram is able to change the number of
crossings of the braid diagram; hence, all braid diagrams in a CPT
multiplet must have the same number of crossings. This exhibits the
latter meaning of the theorem.
\end{proof}

Since in the unique representation, for each number of crossings we
have only four actively-interacting braids, which have been shown
being related only by three discrete transformations, namely
$\mc{C}$, $\mc{P}\mc{T}$ and $\mc{C}\mc{P}\mc{T}$, applying the
remaining four discrete transformations, viz $\mc{P}$, $\mc{T}$,
$\mc{C}\mc{P}$ and $\mc{C}\mc{T}$ can not generate new braids with
the same number of crossings, which means that their actions must be
equivalent to, in certain order, those of $\mc{C}$, $\mc{P}\mc{T}$,
$\mc{C}\mc{P}\mc{T}$, and the identity $\mathds{1}$ on
actively-interacting braids.

As for braids that do not interact actively, we do not have a
similar theorem. In fact, in the unique representation, for
non-actively interacting braids with $m$ crossings ($m>1$), we can
always find those not related to each other by any discrete
transformation. Here is an example with $m=2$: for the braid,
$^{\hspace{0.75mm}S_l}\hspace{-0.5mm}[(T_a,T_b,T_c)\sigma_{ud^{-1}}]^{S_r}$,
and the braid,
$^{\hspace{0.75mm}S'_l}\hspace{-0.5mm}[(T'_a,T'_b,T'_c)\sigma_{uu}]^{S'_r}$,
whatever their internal twists and end-node states are, it is
straightforward to see that they can never be transformed into each
other by the discrete transformations.

Nevertheless, it is still true for non actively-interacting braids
that all braids in a CPT multiplet have the same number of crossings
when they are represented in the same type of representation. This
is so simply because discrete transformations do not change the
representation type and the number of crossings of a braid.
\subsection{Interactions under C, P, and T}
We have seen the effects of C, P, and T on single braid excitations,
it is then natural and important to discuss the action of these
discrete transformations on braid interactions, defined in
\cite{LeeWan2007} graphically, and in \cite{HeWan2008b}
algebraically. Braid interactions turn out to be invariant under
CPT, and more precisely, under C, P, and T separately.

This type of interaction always involve two braids, one of which
must be actively-interacting. As pointed out in \cite{HeWan2008b},
in dealing with an interaction it is convenient to represent the
actively-interacting braid, say $B$, by one of its trivial
representatives, and represent the other braid, say $B'$, by its
unique representative. Although \cite{HeWan2008b} shows that the
right-interaction of $B$ on $B'$, namely $B+B'$, and the
left-interaction $B'+B$ are not equal in general, for the purpose
here we need only to consider either of the two cases because the
other case follows similarly; let us take $B+B'$ to study.

\cite{HeWan2008b} has proven that the interaction $B+B'$ is
independent of the trivial braid diagram representing $B$, and
suggests to choose the one with zero right external twist to
represent $B$. Thus we let
$B=\left._{T_l}^{\hspace{0.75mm}S}\hspace{-0.5mm}[T_a,T_b,T_c]^{S}_{0}\right.$,
and
$B'=\left._{\hspace{1mm}0}^{S_l}\hspace{-0.5mm}[(T'_a,T'_b,T'_c)\sigma_X]^{S_r}_{0}\right.$.
Given this, we can adopt from \cite{HeWan2008b} the algebraic form
of $B+B'$
as follows%
\begin{equation}
B''=B+B'=\left._{T_l}^{S_l}\hspace{-0.5mm}[(T_a+T'_a,T_b+T'_b,T_c+T'_c)\sigma_X]^{S_r}_{0}\right.,
\label{eqBB'}
\end{equation}
where $S=S_l$, such that the so-called interaction condition is
satisfied\cite{LeeWan2007, HeWan2008b}. There is a subtlety, the
braid $B''$ in Eq. \ref{eqBB'} is not the standard result which has
zero external twist and should be obtained from this $B''$ by a
rotation on its left end-node to remove the external twist, $T_l$.
We do so to reduce the complexity of this proof, which has no harm
because rotations, as equivalence moves, obviously commute with
discrete transformations.

We now show that this interaction is invariant under a charge
conjugation, i.e. to show $\mc{C}(B)+ \mc{C}(B')=\mc{C}(B+B')$. By
Tables \ref{tabMap} and \ref{tabDT}, we readily obtain%
\begin{equation}
\begin{aligned}
\mc{C}(B) &=\mc{C}\left(
\left._{T_l}^{\hspace{0.75mm}S}\hspace{-0.5mm}[T_a,T_b,T_c]^{S}_{0}\right.
\right)\\
&=\left._{-T_l}^{\hspace{0.75mm}-S}\hspace{-0.5mm}[-T_a,-T_b,-T_c]^{-S}_{0}\right.\\
\mc{C}(B') &=\mc{C}\left(
\left._{\hspace{1mm}0}^{S_l}\hspace{-0.5mm}[(T'_a,T'_b,T'_c)\sigma_X]^{S_r}_{0}\right.\right)\\
&=\left._{\hspace{3.3mm}0}^{-S_l}\hspace{-0.5mm}[-(T'_a,T'_b,T'_c)\sigma_{\mc{I}_X(X)}]^{S_r}_{0}\right..
\end{aligned}
\end{equation}
Hence, by the same token as how this kind of interaction is carried
out in Eq. \ref{eqBB'}, we get
\begin{align}
\mc{C}(B)+\mc{C}(B') &=
\left._{-T_l}^{\hspace{0.75mm}-S}\hspace{-0.5mm}[-T_a,-T_b,-T_c]^{-S}_{0}\right.
+
\left._{\hspace{3.3mm}0}^{-S_l}\hspace{-0.5mm}[-(T'_a,T'_b,T'_c)\sigma_{\mc{I}_X(X)}]^{S_r}_{0}\right.\nonumber\\
&= \left._{-T_l}^{-S_l}\hspace{-0.5mm}[
-(T_a+T'_a,T_b+T'_b,T_c+T'_c)\sigma_{\mc{I}_X(X)}]^{-S_r}_{0}\right.\label{eqIntWithC}
\end{align}

Now we directly apply a C-transformation on the braid $B''$ in Eq.
\ref{eqBB'}, which leads to
\begin{align}
\mc{C}(B'') &=\mc{C}\left(
\left._{T_l}^{S_l}\hspace{-0.5mm}[(T_a+T'_a,T_b+T'_b,T_c+T'_c)\sigma_X]^{S_r}_{0}\right.\right)\nonumber\\
&=\left._{-T_l}^{-S_l}\hspace{-0.5mm}[
-(T_a+T'_a,T_b+T'_b,T_c+T'_c)\sigma_{\mc{I}_X(X)}]^{-S_r}_{0}\right.,
\end{align}
which is exactly the same as the RHS of Eq. \ref{eqIntWithC}.
Therefore, due to the generality of $B$ and $B'$ all interactions
are invariant under charge conjugation.

We now move on to the case of parity transformation. It is important
to note that a discrete transformation acts on the whole process of
an interaction - in particular the complete states before and after
the interaction. As a result, in view of the fact from Table
\ref{tabDT} that a P-transformation, i.e. the $\mc{F}_H$, exchanges
the two end-nodes, reverses the crossing sequence, and exchanges the
twists on the left and on the right of the crossing sequence of a
braid, it should also switch the positions of the two braids
involved in an interaction before the interaction happens. That is
to say, to show the invariance of an right-interaction, say
$B+B'=B''$ (the same braids as above), under P, one needs to prove
that the left-interaction, $\mc{P}(B')+\mc{P}(B)=\mc{P}(B'')$,
holds\footnote{The interaction condition is automatically satisfied
in this way because the neighboring nodes of B' and B again are in
the same state and have no twist on their common edge, after
switching their positions. Please check \cite{HeWan2008b} for
details on the interaction condition.}.

Applying a P-transformation on $B$ and $B'$ respectively brings us
\begin{equation}
\begin{aligned}
\mc{P}(B) &=\mc{P}\left(
\left._{T_l}^{\hspace{0.75mm}S}\hspace{-0.5mm}[T_a,T_b,T_c]^{S}_{0}\right.
\right)\\
&=\left._{\hspace{2.5mm}0}^{-S}\hspace{-0.5mm}[T_a,T_b,T_c]^{-S}_{T_l}\right.\\
\mc{P}(B') &=\mc{P}\left(
\left._{\hspace{1mm}0}^{S_l}\hspace{-0.5mm}[(T'_a,T'_b,T'_c)\sigma_X]^{S_r}_{0}\right.\right)\\
&=\left._{\hspace{3.5mm}0}^{-S_r}\hspace{-0.5mm}[(T'_{a'},T'_{b'},T'_{c'})\sigma_{\mc{R}(X)}]^{-S_l}_{0}\right.,
\end{aligned}
\end{equation}
where $(T'_a,T'_b,T'_c)\sigma_X=(T'_{a'},T'_{b'},T'_{c'})$. Then
according to \cite{HeWan2008b}, the left interaction of $\mc{P}(B)$
and $\mc{P}(B')$ reads%
\begin{align}
\mc{P}(B')+\mc{P}(B) &=
\left._{\hspace{3.5mm}0}^{-S_r}\hspace{-0.5mm}[(T'_{a'},T'_{b'},T'_{c'})\sigma_{\mc{R}(X)}]^{-S_l}_{0}\right.
+
\left._{\hspace{2.5mm}0}^{-S}\hspace{-0.5mm}[T_a,T_b,T_c]^{-S}_{T_l}\right.\nonumber\\
&=\left._{\hspace{3.5mm}0}^{-S_r}\hspace{-0.5mm}[((T'_{a'},T'_{b'},T'_{c'})+
\sigma^{-1}_{\mc{R}(X)}(T_a,T_b,T_c))\sigma_{\mc{R}(X)}]^{-S_l}_{T_l}\right.\nonumber\\
\xRightarrow{Eq. \ref{eqSigmaRinverseSigmaX}}
&=\left._{\hspace{3.5mm}0}^{-S_r}\hspace{-0.5mm}[((T'_a,T'_b,T'_c)\sigma_X+
(T_a,T_b,T_c)\sigma_X)\sigma_{\mc{R}(X)}]^{-S_l}_{T_l}\right.\nonumber\\
&=\left._{\hspace{3.5mm}0}^{-S_r}\hspace{-0.5mm}[((T'_a+T_a,T'_b+T_b,T'_c+T_c)\sigma_X)
\sigma_{\mc{R}(X)}]^{-S_l}_{T_l}\right..\label{eqIntWithP}
\end{align}
Note that in the last line of the equation above the $\sigma_X$ and
$\sigma_{\mc{R}(X)}$ should not be contracted by Eq.
\ref{eqSigmaSigmaRx} and we did not do so, because the
$\sigma_{\mc{R}(X)}$ is present there not only for denoting the
permutation but also for recording the crossing sequence, namely
$\mc{R}(X)$, of the resulted braid.

If we directly apply a P-transformation on $B''$ in Eq. \ref{eqBB'},
we attain%
\begin{equation*}
\begin{aligned}
\mc{P}(B'') &=
\mc{P}\left(\left._{T_l}^{S_l}\hspace{-0.5mm}[(T_a+T'_a,T_b+T'_b,T_c+T'_c)\sigma_X]^{S_r}_{0}\right.
\right)\\
&=\left._{\hspace{3.5mm}0}^{-S_r}\hspace{-0.5mm}[((T'_a+T_a,T'_b+T_b,T'_c+T_c)\sigma_X)
\sigma_{\mc{R}(X)}]^{-S_l}_{T_l}\right.,
\end{aligned}
\end{equation*}
for that the triple $(T'_a+T_a,T'_b+T_b,T'_c+T_c)\sigma_X$ is
already the original triple of internal twists on the right of $X$,
which should become the left one of $\mc{P}(B'')$ due to the effect
of parity transformation. This equation is precisely the one on the
RHS of Eq. \ref{eqIntWithP}. Therefore, the invariance of braid
interaction under parity transformation is established.

Regarding interactions under time reversal there is a subtlety. The
time reversal we have found is with respect to a single braid
excitation; however, an interaction involves the time evolution of a
spin network under dynamical moves. To apply our T-transformation to
an interaction, one should reverse all the dynamical moves, e.g. a
$1\rightarrow 4$ move taken in the interaction becomes a
$4\rightarrow 1$ move under the time reversal, at the same time. In
this sense, to show the invariance of the interaction,
$B+B'\longrightarrow B''$, it suffices to show that
$\mc{T}(B'')\longrightarrow \mc{T}(B')+\mc{T}(B)$, where the
positions of $B$ and $B'$ are swapped as in the case of parity
transformation. It is straightforward to prove the invariance of
interactions under time reversal; the procedure is similar to that
of the previous two cases, and is thus not explicitly presented
here.

Consequently, all braid interactions of the type defined in
\cite{LeeWan2007, HeWan2008b} are invariant under C, P, T, and any
combination of them. This is not the case of the Standard Model of
particles because of the absence of CP-violation. A possible reason
is that all our current interactions are deterministic. However, it
is suggested in \cite{HeWan2008b} that one may consider
superpositions of braids by possibly taking spin network labels into
account. This is to be discussed in the next section.

It is however necessary to remark that the current study of discrete
transformations of braids would not be impact by just adding spin
network labels in a straightforward way in to our scheme. A reason
is that the discrete transformations found in this paper do not
change the spin network label of each existing edge of the network.
One may try to construct discrete transformation of braids which
change the spin network labels, but there could be many arbitrary
ways of doing this and no a priori reason of making a special
choice.
\section{Discussion and conclusion}
The results we have obtained so far are purely based on the algebra
of the action of discrete transformations on topological quantities
characterizing braids. Although our braids live on embedded spin
networks, spin network labels have yet not been incorporated along
this research line; spin networks are treated as framed graphs with
merely topological properties. However, for reasons which will be
clear soon, there is a necessity for spin network labels to be
included.

The mapping between the discrete transformations of braids and the
C, P, and T were determined without referring to the definition of
the spin of a braid. This is because our analysis indicates that
spin cannot be constructed out of the conserved topological
quantities we have in hand of braids. The reasons are as follows.

In our language crossings of a braid and twists on the strands of
the braid are on an equal footing due to equivalence moves which can
trade crossings with twists or vice versa, which leads to the
effective twist number $\Theta$ of a braid, a conserved quantity
independent of the representative of the braid. For the
aforementioned reason we may identify $\Theta$ or certain
appropriate function of it as the charge of a braid. There is no a
consistent way to directly associate $\Theta$ with spin as well.
Furthermore, since in the context of particle physics, charge is a
result of $U(1)$ gauge symmetry (and our twists are also related to
$U(1)$ as previously explained), while spin results from
Poincar\'{e} symmetry, a space-time symmetry, charge and spin thus
could not be unified by superficially manipulating $\Theta$.

Spin network labels are usually representations of a gauge group or
of a quantum group. In both of the original spin network
proposal\cite{Penrose} and the Loop Quantum Gravity, spin network
labels are $SU(2)$ or $SO(3)$ representations. In the framed case,
one uses, for example, the quantum $SU(2)$, namely $SU_q(2)$.
Consequently, it is more reasonable to associate the spin of a braid
with the spin network labels on the strands and/or external edges,
and the intertwiners at the end-nodes of the braid. If one decorate
the 4-valent spin networks by representations of gauge groups other
than $SU(2)$ and $SO(3)$, say $SU(3)$, it may be possible to define
color charges of braids as well.

If we agree on this, then Table \ref{tabDTcq} gives us a hint how we
would find spin for braids. The time reversal is identified with the
discrete transformation $\mc{F}_V\mc{F}_H$. Since a T-transformation
flips the $z$-component of the spin of a single-particle state, and
because external twists and internal twists of a braid are not
separately conserved quantities independent of representatives of
the braid, by the last line of Table \ref{tabDTcq}, the only effect
of $\mc{F}_V\mc{F}_H$ on a braid, which possibly corresponds to the
change of spin due to time reversal, is then the exchange of the top
and the bottom strands of the braid, which is implied by the
exchange of the top and the bottom internal twists explicitly shown
in the table. This implies that the spin network labels respectively
on the top and the bottom strands are also subject to exchange under
a $\mc{F}_V\mc{F}_H$, or simply a T-transformation. In other words,
we would like to have a way of combining labels and/or intertwiners
of a braid, which changes sign when the label on the top strand and
the one on the bottom strand of the braid are exchanged. The sign
factor of the action of any discrete transformation involving a time
reversal, and other phase factors, in Table \ref{tabCPT} will appear
accordingly. Unfortunately, the precise form of this construction is
unavailable for the moment; its discovery may require a complete and
systematic approach which takes spin network labels into account.

Our observation that all interactions of braids are invariant under
C, P, and T separately seems to indicate an issue that the
interactions of braids we have studied so far are deterministic, in
the sense that an interaction of two braids produces a definite new
braid. Nevertheless, this may not be a problem at all because it
could be due to the simple fact that we have only worked with
definite vertices of interactions. In terms of vertices we have
definite result for an interaction as to the case in QFT; this is
similar to what have been done in spin foam models or group field
theories. Besides, one can certainly argue that if our braids are
more fundamental entities, the CP-violation in particle physics does
not necessarily exist at this level. Putting this CP-violation
problem aside; however, a fully quantum mechanical picture should be
probabilistic\footnote{One should note that a few theoretical
physicists may not agree on this.}.

If we tend to take this as an issue, we may try to work with
superpositions of braids and interactions of braids resulting in
superposition of braids. A possible way out is to consider braids
with the same topological structure but different sets of spin
network labels as physically different. One may adopt from spin foam
models the methods which can assign amplitudes to the dynamical
moves, namely the dual Pachner moves restricted by the stability
condition\cite{LeeWan2007, Isabeau2008}, of the embedded 4-valent
spin networks. A dynamical move such as a $2\rightarrow 3$ and a
$1\rightarrow 4$ may then give rise to outcomes with the same
topological configuration but different spin network labels; each
outcome has a certain probability amplitude. As a result, an
interaction of two braids may give rise to superposed braids, each
of which has a certain probability to be observed, with the same
topological content but different set of spin network labels. With
this, CP-violating interactions may arise.

There seems to be a more elegant and unified way to resolve all the
aforementioned issues once for all, which is the so-called tensor
category, or more specifically the braided tensor category with
twists. As previously explained, a twist of a strand of a braid can
be interpreted as characterizing a non-trivial isomorphism from
$U(1)$ to $U(1)$. However, the concept of twist can be generalized
to vector spaces other than representation spaces of $U(1)$. This is
the way how twists are defined in the language of tensor categories.
In this manner, we may view spin network labels as they represent
generalized framing of spin networks other than the $U(1)$ framing
we have just studied, such that generalized twists can arise. The
consequence is our twists and spin network labels, and hence gauge
symmetries and space-time symmetries may be unified in this way.

Tensor categories naturally use isomorphisms between tensor products
of vector spaces to account for braiding. This can be understood,
for example, from the solutions of (Quantum) Yang-Baxter Equations.
However, it is important to note that our braids are special because
each of them has two 4-valent end-nodes and two external edges. All
these will exert further constraints on the possible tensor
categories we can use, or motivate new types of tensor categories.
Tensor-categorized 4-valent braids and evolution moves may be
evaluated by the relevant techniques already defined in theories of
tensor category or new techniques adapted to our case.

There are also other works on unification of gravity and matter or
on emergent matter degrees of freedom, which indicate that tensor
category might be a correct underlying mathematical language towards
this goal. The string network condensation by Wen \emph{et
al}\cite{WenXiaogang} is such an example.

In conclusion, we have found seven discrete transformations of
3-strand braids and mapped them to C, P, T, and their products.
Along with this, the effective twist number of a braid has been
demonstrated to be responsible for the electric charge of the braid.
It is very interesting that in the braid representation without
external twist, all actively-interacting braids of the same number
of crossings form a CPT multiplet, whereas there are non
actively-interacting braids of the same number of crossings which
can not be transformed into each other by any legal discrete
transformation of our braids. This will help us to find a deeper
correspondence between our braids and matter particles. Furthermore,
braid interactions have been proven invariant under C, P, and T
separately.

We have explained the necessity to incorporate spin network labels
into this approach. This allows us to argue that the spin of a braid
is related to the spin network labels of the braid. In addition,
probability amplitudes of braid propagation and interaction may also
be constructed with the help of spin network labels. A possible
future direction regarding a more generalized approach, in terms of
tensor categories, along this research line is pointed out.
\section*{Acknowledgements}
We thank Lee Smolin, Gerard 't Hooft, and Donald Marolf for helpful
discussions and comments. We strongly appreciate Yong-Shi Wu for
motivating a very interesting future direction along this research
line. YW's gratitude also goes to Catherine Meusburger and Jonathan
Hackett for discussions. SH is also grateful to the hospitality of
the Perimeter Institute. Research at the Perimeter Institute for
Theoretical Physics is supported in part by the Government of Canada
through NSERC and by the Province of Ontario through MRI. Research
at Peking University is supported by NSFC (nos.10235040 and
10421003).

\section*{Appendix I: Atomic discrete operations on braids}
\addcontentsline{toc}{section}{Appendix I: Atomic discrete
operations on braids}%
It is useful to represent a legal discrete transformation in an
algebraic form. However, a discrete transformation of a braid
normally acts on all elements in the characterizing 8-tuple of the
braid. Consequently, one can split an action of a discrete
transformation to minimal sub-operations on components of the
8-tuple. These minimal sub-transformations are named \textbf{atomic
discrete transformation}s, or atomic operations for short. An atomic
operation can only act on one and only one type of component in the
characterizing 8-tuple of a braid because otherwise it can be
divided again and hence is not atomic.

Let us be precise. The characterizing 8-tuple of a braid, say
$\{T_l,S_l,T_a,T_b,T_c,X,S_r,T_r\}$, consists of four types of
components-to wit $(T_l, T_r)$, the pair of external twists, $(T_a,
T_b,T_c)$, the triple of internal twists, the crossing sequence $X$,
and the pair of end-node states, $(S_l,S_r)$. In some cases, the
external twists and internal twists can be considered together due
to the fact that they are transformed simultaneously in the same
manner. An atomic operation is only allowed to act on one of these
four types or on the set of all twists. In addition, if an atomic
operation acts on an element in the 8-tuple it must also acts on all
other elements of the same type in a similar way because otherwise
Condition \ref{condLegalDT} would be violated. This will be
clarified case by case. We now try to sort out all legal atomic
operations.

Since twists, crossings and end-node states can take both positive
and negative values in our framework, it is natural to have discrete
operations which flip their values. There are three such atomic
operations, called inversions.
\begin{definition}\label{defIs}
The \textbf{inversion of the end-node states} of a braid, denoted by
$\mc{I}_S$, is an atomic operation flipping the signs of both
end-node states of the braid. That is,$$\mc{I}_S:\
(S_l,S_r)\mapsto(\bar{S_l},\bar{S_r}).$$
\end{definition}
\begin{definition}\label{defIx}
The \textbf{inversion of the crossing sequence} $X$ of a braid,
denoted by $\mc{I}_X$, is an atomic operation taking each crossing
in $X$ to
its inverse, namely%
\beqq
\begin{aligned}
\mc{I}_X:\ &u\mapsto u^{-1}\\
           &d\mapsto d^{-1}\\
           &X=x_1x_2\cdots x_n\mapsto x^{-1}_1x^{-1}_2\cdots
           x^{-1}_n.
\end{aligned}
\eeqq
\end{definition}
The integral value of each crossing is negated by this operation, so
is the crossing number of the braid. In addition, we clearly have
$\sigma_X=\sigma_{\mc{I}_X}$. The two atomic operations above must
act on both end-nodes and on all crossings of $X$ respectively.
Otherwise, one cannot combine them to make a discrete transformation
which is legal for arbitrary braids.
\begin{definition}\label{defIt}
The \textbf{inversion of the twists} of a braid, denoted by
$\mc{I}_T$, is an atomic operation which multiplies a $-1$ to every
twist of the braid, i.e. $$\mc{I}_T:\ \{T_l,T_a,T_b,T_c,T_r\}\mapsto
\{-T_l,-T_a,-T_b,-T_c,-T_r\}.$$
\end{definition}
That this atomic discrete transformation acts on all the twists,
internal and external, of a braid is largely due to Condition
\ref{condLegalDT}. The reason is that if not all but one or several
of its twists are flipped, there is no way to combine such an
operation with other atomic operations, albeit all other operations
are legal, to keep any actively-interacting braids active, since
twists play a key role in the activity of a braid.
There are two more atomic operations acting on $X$.%
\begin{definition}\label{defR}
The \textbf{reversion} is an atomic operation, denoted by $\mc{R}$,
which reverses the order of the crossings in a crossing sequence
$X$:
$$\mc{R}:\ X=x_1x_2\cdots x_n\mapsto x_n x_{n-1}\cdots x_1.$$
\end{definition}
It is also useful for the sake of calculation to define $X^{-1}$ to
be the combined result of $\mc{I}_X$ and $\mc{R}$ on $X$, i.e.
$X^{-1}=\mc{I}_X\mc{R}(X)$. Note that for the permutation induced by
$X$, $\sigma^{-1}_X\neq\sigma_{X^{-1}}$ in general. However, it is
quite clear that%
\begin{equation}
\sigma_X\sigma_{\mc{R}(X)}=\sigma^{-1}_{\mc{R}(X)}\sigma^{-1}_X\equiv\mathds{1}%
\label{eqSigmaSigmaRx}
\end{equation}
The meaning of this equation must be understood from its action on
triples of internal twists. That is,
$$(T_a,T_b,T_c)\sigma_X\sigma_{\mc{R}(X)}=(T_a,T_b,T_c).$$ Keeping in mind that
$\sigma^{-1}_X$ is a left-acting function, if we apply a
$\sigma^{-1}_X$ to the left of both sides of the above
equation, we get%
\begin{equation}
\begin{aligned}
\sigma^{-1}_X(T_a,T_b,T_c)\sigma_X\sigma_{\mc{R}(X)}
&= \sigma^{-1}_X(T_a,T_b,T_c)\\
\xRightarrow{Eq. \ref{eqSigmaTsigma}}
(T_a,T_b,T_c)\sigma_{\mc{R}(X)}&=\sigma^{-1}_X(T_a,T_b,T_c).
\end{aligned}
\label{eqSigmaRsigmaXinverse}
\end{equation}
Similary, what follows is also true:%
\begin{equation}
\sigma^{-1}_{\mc{R}(X)}(T_a,T_b,T_c)=(T_a,T_b,T_c)\sigma_X.%
\label{eqSigmaRinverseSigmaX}
\end{equation}

\begin{definition}\label{defSc}
A \textbf{chain shift} is an atomic operation on $X$, denoted by
$\mc{S}_c$, shifting every upper crossing in $X$ to a lower one and
a lower one to an upper one, with however, the crossing's chirality
intact. That
is,%
\beqq \mc{S}_c:\ \forall x_i\in X, x_i\mapsto%
\begin{cases}
d,\ \mathrm{if}\ x_i=u\\
u,\ \mathrm{if}\ x_i=d%
\end{cases}
,\ i=1\dots n.%
\eeqq
\end{definition}
The above two atomic operations on $X$ must apply to every crossing
in $X$ simultaneously because otherwise an alternating braid could
be transformed into a non-alternating braid and vice versa, which
certainly causes the violation of Condition \ref{condLegalDT} if
they are part of a discrete transformation of a braid.

There is actually a hidden triple in the characterizing 8-tuple of a
braid, i.e. the triple of internal twists on the right of the
crossing sequence $X$, $(T_{a'},T_{b'},T_{c'})$. It is not
explicitly included in the 8-tuple because is is related to the
triple $(T_a,T_b,T_c)$ by the induced permutation $\sigma_X$, as
aforementioned. However, one can have a transformation which
exchanges these two triples.%
\begin{definition}\label{defEt}
The \textbf{exchange of triples of internal twists} of a braid is an
atomic operation, denoted by $\mc{E}_T$, doing the following:
$$\mc{E}_T:\ (T_a,T_b,T_c)\mapsto(T_{a'},T_{b'},T_{c'})$$
and vice versa, where
$(T_a,T_b,T_c)\sigma_X=(T_{a'},T_{b'},T_{c'})$.
\end{definition}
An exchange of triples of internal twists is usually accompanied by
an exchange of the two external twists of a braid.
\begin{definition}\label{defEte}
The \textbf{exchange of external twists} of a braid, $\mc{E}_{T_e}$,
is an atomic operation, such that
$$\mc{E}_{T_e}:\ (T_l,T_r)\mapsto(T_r,T_l).$$
\end{definition}

The last possible atomic discrete transformation on the twists is
defined as follows.
\begin{definition}\label{defSt}
The \textbf{twist swap} is an atomic operation, denoted by
$\mc{S}_T$, which swaps the top and bottom internal twists of a
braid:
$$\mc{S}_T:\ (T_a,T_b,T_c)\mapsto(T_c,T_b,T_a).$$
\end{definition}
Finally, one can exchanges the two end-node states of a braid.
\begin{definition}\label{defEs}
The exchange of end-node states, $\mc{E}_S$, is an atomic operation,
such that
$$\mc{E}_S:\ (S_l,S_r)\mapsto(S_r,S_l).$$
\end{definition}

An important remark is that all above atomic operations in fact act
on braids. For simplicity nevertheless, we only show here their
definitions their effects on the relevant characterizing quantities
of a braid. Another remark is that all atomic operations commute
with each other and hence their relative positions in a combination
as a discrete transformation do not matter.

\section*{Appendix II: Examples of braids under C, P, and T}
\addcontentsline{toc}{section}{Appendix II: Examples of braids under C, P, and T}%

\begin{equation*}
\begin{aligned}
                           & \raisebox{-3.82ex}{{\includegraphics[
height=0.6754in, width=3.717in
]%
{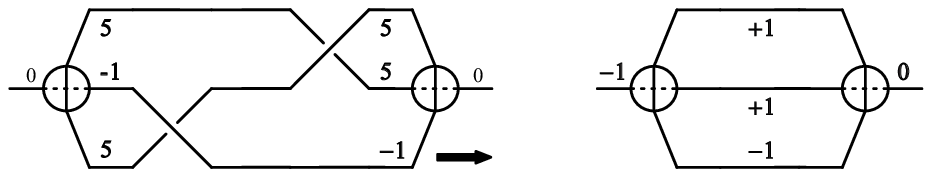}%
}}\\%
\xRightarrow{\mc{M}_{\Box}} \hspace{15mm} &
\raisebox{-3.82ex}{{\includegraphics[ height=0.6754in, width=3.717in
]%
{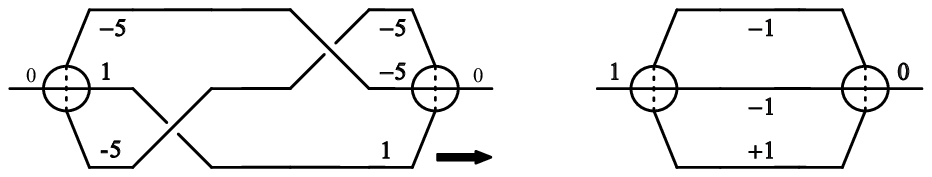}%
}}\\%
\xRightarrow{\mc{F}_H} \hspace{15mm} &
\raisebox{-3.82ex}{{\includegraphics[ height=0.6754in, width=3.717in
]%
{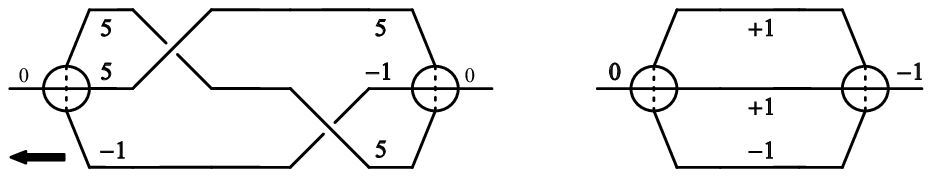}%
}}
\\%
\xRightarrow{\mc{F}_V\mc{F}_H} \hspace{15mm} &
\raisebox{-3.82ex}{{\includegraphics[ height=0.6754in, width=3.717in
]%
{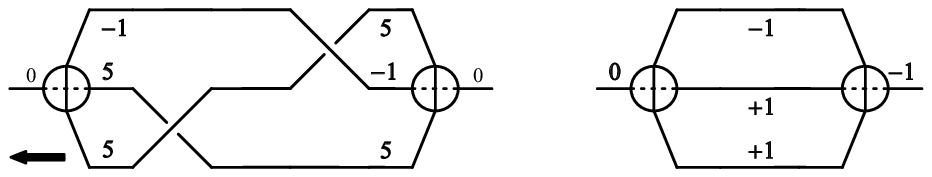}%
}}
\\%
\xRightarrow{\mc{M}_{\perp}} \hspace{15mm} &
\raisebox{-3.82ex}{{\includegraphics[ height=0.6754in, width=3.717in
]%
{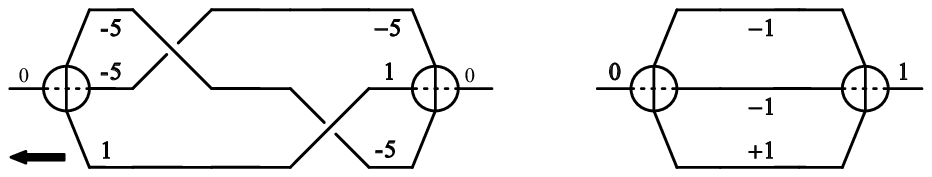}%
}}
\\%
\xRightarrow{\mc{M}_{\perp}\mc{F}_V} \hspace{15mm} &
\raisebox{-3.82ex}{{\includegraphics[ height=0.6754in, width=3.717in
]%
{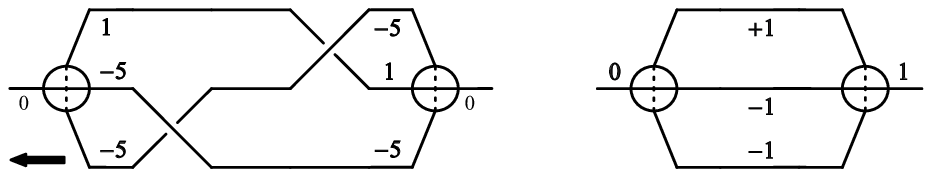}%
}}
\\%
\xRightarrow{\mc{F}_V} \hspace{15mm} &
\raisebox{-3.82ex}{{\includegraphics[ height=0.6754in, width=3.717in
]%
{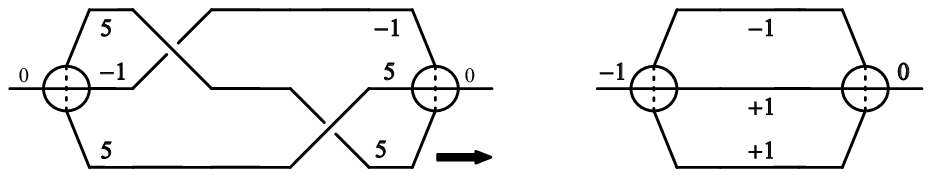}%
}}
\\%
\xRightarrow{\mc{M}_{\Box}\mc{F}_V} \hspace{15mm} &
\raisebox{-3.82ex}{{\includegraphics[ height=0.6754in, width=3.717in
]%
{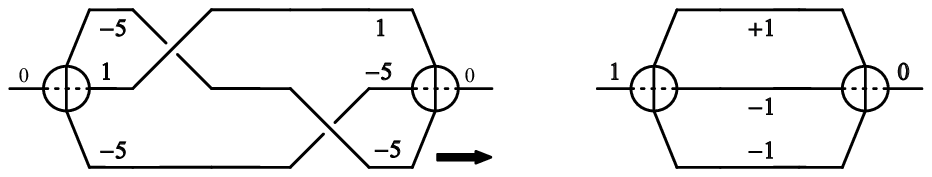}%
}}
\end{aligned}
\end{equation*}

The figure above illustrates two examples of braids under the action
of $G_{\mc{D}}$. The first row shows the two original braid
diagrams: the left one is a right propagating braid in zero external
twist representation, while the right one is an actively-interacting
braid represented by a trivial braid diagram. The thick arrow on the
lower left or right corner of a braid diagram indicates the
propagation chirality of the braid.


\end{document}